\documentclass[sigconf,balance=false,authorversion,nonacm]{acmart}
\usepackage{popets}
\usepackage{comment}
\usepackage{tabularx}
\usepackage{booktabs}
\usepackage{breakcites}

\usepackage{tikz}
\newcommand*\emptycirc[1][1ex]{\tikz\draw (0,0) circle (#1);} 

\newcommand*\fullcirc[1][1ex]{\tikz\fill (0,0) circle (#1);}

\setcopyright{popets}
\copyrightyear{2025}

\acmYear{2025}
\acmVolume{2025}
\acmNumber{X}
\acmDOI{XXXXXXX.XXXXXXX}
\acmISBN{}
\acmConference{Proceedings on Privacy Enhancing Technologies}
\begin{document}

\title[SoK: Web Authentication in the Age of End-to-End Encryption]{SoK: Web Authentication in the Age of\\ End-to-End Encryption}


\author{Jenny Blessing}
\affiliation{%
  \institution{University of Cambridge}
  \country{}}
\email{jenny.blessing@cl.cam.ac.uk}

\author{Daniel Hugenroth}
\affiliation{%
  \institution{University of Cambridge}
  \country{}}
\email{daniel.hugenroth@cl.cam.ac.uk}

\author{Ross J. Anderson}
\affiliation{%
  \institution{University of Cambridge \\ University of Edinburgh}
  \country{}
}

\author{Alastair R. Beresford}
\affiliation{%
 \institution{University of Cambridge}
 \country{}}
\email{alastair.beresford@cl.cam.ac.uk}



\renewcommand{\shortauthors}{Blessing et al.}

\begin{abstract}
The advent of end-to-end encrypted (E2EE) messaging and backup services has brought new challenges for usable authentication.
Compared to regular web services, the nature of E2EE implies that the provider cannot recover data for users who have forgotten passwords or lost devices.
Therefore, new forms of robustness and recoverability are required, leading to a plethora of solutions ranging from randomly-generated recovery codes to threshold-based social verification.
These implications also spread to new forms of authentication and legacy web services: passwordless authentication (``passkeys'') has become a promising candidate to replace passwords altogether, but are inherently device-bound.
However, users expect that they can login from multiple devices and recover their passwords in case of device loss---prompting providers to sync credentials to cloud storage using E2EE, resulting in the very same authentication challenges of regular E2EE services.
Hence, E2EE authentication quickly becomes relevant not only for a niche group of dedicated E2EE enthusiasts but for the general public using the passwordless authentication techniques promoted by their device vendors.
In this paper we systematize existing research literature and industry practice relating to security, privacy, usability, and recoverability of E2EE authentication.
We investigate authentication and recovery schemes in all widely-used E2EE web services and survey passwordless authentication deployment in the top-200 most popular websites.
Finally, we present concrete research directions based on observed gaps between industry deployment and academic literature.
\end{abstract}


\maketitle

\section{Introduction}
Password-based threats have plagued web authentication for as long as the Internet has existed, with Google declaring passwords to be ``the single biggest threat to your online security’’ in 2021~\cite{risher_blog21}. 
While industry providers have deployed a wide variety of end-user authentication and recovery schemes in the past decade, developing an authentication scheme that offers a desirable balance of security, privacy, usability, and recoverability for a diverse population of users is an enormous challenge. 
The inherent difficulty of resolving these trade-offs is the central reason why passwords persist~\cite{herley2011research,quermann2018state,bonneau2012quest}. 
Even so, industry providers have made significant strides in mitigating password-based threats, such as using browser metadata to detect suspicious logins~\cite{wiefling2019really,freeman2016you,milka2018anatomy} and offering ``single sign-on'' options to centralize authentication with a single account.

We are entering a new era: the FIDO2 standard~\cite{fido2_protocol} (jointly developed by the World Wide Web Consortium (W3C) and the FIDO Alliance) provides a password-less authentication protocol based on public-key cryptography, commonly referred to as \emph{passkeys}.
Passkeys have recently been deployed across all major operating systems and web browsers, and as of May 2024 over 400 million Google accounts have set up a passkey~\cite{google_passkeys_blog24}.
End-users are able to authenticate to remote web servers using their device authentication (e.g., fingerprint, PIN, etc.) to unlock access to the relevant key material. 
While passwords as a whole will not disappear any time soon since they remain the most usable authentication scheme for many, especially at-risk demographics~\cite{ahmed2019everyone,sambasivan2018privacy}, passkeys for the first time provide a genuine contender for password replacement which has the backing of major industry players. 
However, there are still numerous unresolved questions around usability, security, and recoverability, all of which will have a major impact on end-user adoption in practice. 
To enable recoverability in the case of device loss, deployed industry solutions offer users the option to sync device passkeys to the device's cloud provider, where the stored keys are end-to-end encrypted (typically using the device passcode) such that they are inaccessible to the provider.

Another significant development in the authentication landscape is that a small but notable number of consumer-facing cloud storage and email providers have begun protecting user data such as photos, videos and documents with end-to-end encryption (E2EE).
This marks an important shift, as E2EE has historically been limited to securing ephemeral communications, or particularly sensitive data such as passwords.
This development raises concerns around data recoverability since, by definition, service providers cannot access user data in an E2EE service. 
Therefore users must take an active role in maintaining access to their account, often including the responsibility for storing their decryption keys.
Providers offer a variety of recovery mechanisms to prevent users losing access to their data, some more user-friendly than others, but there is an unavoidable trade-off in any discussion of user account security, E2EE or otherwise: the easier it is for a user to recover their account, the easier it is for an attacker to gain access. 
At the same time, loss of cloud storage containing years of photo and documentation storage is a much more concerning prospect than loss of ephemeral communication history for most users, motivating a fresh look at industry deployments.

Taken together, the combination of E2EE credential storage and E2EE services represent a rapid shift towards deploying E2EE within web authentication and mobile apps. 
While some E2EE services are generally only adopted by a niche segment of users, passwordless authentication is targeted at the general public. 
Increasingly, though an individual website may not adopt E2EE in any capacity, the credentials of a percentage of the websites' end users will be E2EE, raising the prospect of inadvertent loss of access.

In this work, we systematize contemporary authentication and recovery mechanisms for web applications, with a particular focus on E2EE credentials. 
We analyze both academic literature and industry deployments, survey the usage of these schemes in the most popular websites, and identify challenges to deployment at scale. 
End-user authentication is at an inflection point, as companies are in the process of deploying novel security and authentication schemes representing a significant departure from existing user experiences. 
This is the ideal moment to take stock of benefits and drawbacks offered by newer schemes compared to older ones, and review lessons learned from historical schemes to avoid making similar mistakes.

We begin by synthesizing existing research on the five primary authentication types we observe widely discussed in academic literature and deployed in industry and discuss relevant security, privacy, and usability considerations. 
We only consider user-facing authentication schemes—e.g., the credentials provided by the user to identify themselves with a particular account, rather than client-to-server authentication protocols used to secure credentials. 
We focus particularly on recovery mechanisms, as these are often less secure than the primary authentication mechanism and consequently the means by which an adversary is able to gain access~\cite{hill2017moving,gelernter2017password}.

We conduct the first survey of passwordless authentication adoption among the most widely used websites within the U.S as of May 2024 to better understand which schemes are deployed and where. 
We further conduct an in-depth review of authentication and recovery processes for E2EE web services (including E2EE cloud storage, email, and messaging), finding that authentication methods in this context vary widely. 
Many service providers rely on asking the user to manually store a recovery key even as more usable variations are feasible, a design choice that arguably makes end users less likely to enable E2EE backups and ultimately undermines the very significant privacy enhancement E2EE provides. 
We believe that some authentication factors that may be overly cumbersome as part of a daily authentication scheme, such as authentication using trusted contacts, are worth revisiting in the context of E2EE recovery.

Our specific contributions are the following:

\begin{itemize}
    \item Systematize existing research literature and industry practice relating to security, privacy, usability, and recoverability of all major categories of end-user authentication, with a particular focus on E2EE authentication and recovery.
    \item Investigate currently deployed authentication and recovery schemes in E2EE web services and survey the state of passwordless authentication deployment.
    \item Discuss key trends in the contemporary authentication landscape and provide concrete research directions based on observed gaps between industry deployment and academic literature.
\end{itemize}

\section{Related Work}

\subsection{Authentication Frameworks}
In 2012 Bonneau et al.~\cite{bonneau2012quest} introduced a seminal framework for evaluating authentication schemes across security, privacy, and usability, concluding at the time that no scheme met as many desired criteria as conventional password-based authentication. While the desirable properties of any authentication mechanism deployed at scale have generally remained the same, both state-of-the-art industry deployments and academic research understanding have evolved dramatically since their survey---device-based credential protocols such as FIDO2 did not yet exist, and industry had only just begun introducing the concept of multi-factor authentication (MFA)~\cite{grosse2012authentication}. In 2017, Alomar et al.~\cite{alomar2017social} presented a framework for classifying social authentication schemes and associated attacks, though in practice few of the social authentication schemes are deployed outside of trusted contact recovery. In 2021, Kunke et al.~\cite{kunke2021evaluation} evaluated 12 common account recovery mechanisms within the 2012 framework of Bonneau et al., though at the time the only FIDO2 protocol deployments were hardware token-based and thus they did not consider passwordless authentication.

Prior comparative surveys of authentication schemes were either conducted well before major contemporary shifts in authentication schemes (namely, MFA adoption, device-based authentication, and E2EE authentication) or focused on authentication schemes pre-dating recent trends (as in Lassak et al.~\cite{lassak2024comparative}, a longitudinal study of the usability of email, SMS, recovery questions, and social authentication as fallback mechanisms). To the best of our knowledge no prior work has surveyed authentication and recovery schemes in an end-to-end encrypted context.




\subsection{Measuring Industry Deployments}
A handful prior studies have surveyed industry deployments of non-E2EE authentication schemes. In a 2021 survey of the 208 most widely used websites that offer account creation, Gavazzi et al.~\cite{gavazzi2023study} found that only 42.3\% of accounts support MFA and approximately 22\% appear to support some form of risk-based authentication (e.g., blocking suspicious login attempts based on geolocation). In 2019, Ulqinaku et al.~\cite{ulqinaku2021real} found that 23 of the Alex top 100 websites support the Universal 2nd Factor (U2F) hardware token protocol, but did not find any sites supporting passwordless authentication at the time. More recently, in 2023 Kuchhal et al.~\cite{kuchhal2023evaluating} surveyed the prevalence of the Web Authentication (WebAuthn) API used to provide public-key authentication (and deployed as part of various MFA schemes and passwordless authentication), finding that while 85 of the 585 domains in the Tranco Top 1K~\cite{pochat2018tranco} that supported account creation also supported the WebAuthn protocol, the vast majority used it to support some form of MFA, not passwordless authentication.

In this work, we particularly investigate end-to-end encrypted web services, where the account which the user is attempting to recover is encrypted such that the provider definitionally can be of no help, and furthermore the user may no longer have access to their password or other primary authentication mechanism (e.g., mobile device). Holtervennhoff et al.~\cite{holtervennhoff2024mixed} recently conducted a usability survey of users' perceptions and strategies for handling E2EE recovery keys, but no prior work has looked comprehensively at deployed authentication and recovery mechanisms for end-to-end encrypted data, where either the web service or credentials may be E2EE.

\section{The State of Web Authentication and Recovery}
\label{sec:the-state-of-web-authentication-and-recovery}

We broadly group contemporary authentication schemes into one of two categories, \textit{primary} and \textit{secondary}, based on the context in which it is most widely deployed. Primary authentication mechanisms are the first (and often the only) step required for identity verification, while secondary authentication mechanisms are usually only triggered after the primary authentication process finishes correctly. Both primary and secondary authentication mechanisms are not mutually exclusive in real deployments--providers often allow users to choose among multiple secondary authentication mechanisms, or use one secondary factor as a fallback for another (e.g., using recovery codes in case of MFA app loss).

For the purposes of this section we do not distinguish between authentication and recovery schemes since our goal is to summarize pathways towards account access. From an attacker's perspective, there is no distinction between authentication and recovery. While we focus mainly on schemes that have been widely deployed, we include further discussion of schemes proposed in academic work but that have not gained traction at scale in Appendix~\ref{sec:appendix-auth}.

\subsection{Primary Authentication Mechanisms}

\subsubsection{Passwords}
Security and usability issues with passwords are legion~\cite{florencio2007strong,florencio2007large,herley2009passwords,thomas2017data,nisenoff2023two,bonneau2010password}. Decades of academic research has shown that users constantly forget passwords~\cite{bonneau2015secrets,stobert2018password}, use guessable passwords~\cite{wang2016targeted,bonneau2012science,pal2019beyond}, reuse passwords across different accounts and providers as a coping mechanism~\cite{das2014tangled,wash2016understanding,abbott2018factors,seitz2017differences,inglesant2010true,florencio2014password}, and opt not to change their password even when notified of password reuse or insecurity~\cite{golla2018site,wang2018end}. In the contemporary threat landscape, however, even widely recommended security practices such as increasing password strength would do little protect against phishing attacks~\cite{thomas2019protecting,florencio2007strong} and large-scale data breaches.~\cite{risher_blog21,milka2018anatomy} even if users were to adopt best practices. These concerns impact users at all skill levels—academic work has found the relationship between technical expertise and vulnerability to common attacks (including susceptibility to phishing attacks, password reuse, and choosing stronger passwords) is largely inconclusive~\cite{wei2024sok,lain2022phishing}.

In addition to public databases allowing users to check whether their credentials have been compromised, industry has deployed various cryptographic techniques to automatically alert users of password reuse and breach, such as Meta’s Private Data Lookup (PDL) tool using private set intersection to check whether a user’s password is contained within a server-side set of passwords exposed in data breaches~\cite{meta_pdl23}. Unfortunately, academic work has repeatedly shown that the effectiveness of user notifications is limited: Only around a quarter of warnings resulted in users changing their password~\cite{thomas2019protecting,golla2018site}. Given the unavoidable tensions between security and usability in any password-based authentication scheme, passwords are increasingly viewed as a ``legacy authentication mechanism''~\cite{google_passkeys24}.

\vspace{0.5em}
\noindent\textbf{Password Managers:} A mitigation strategy to make it easier for users to handle vast quantities of credentials is to recommend users use a password manager. While the technical community favors password managers for security reasons as they allow users to opt for higher-entropy passwords and eliminates the need to memorize credentials, academic work has shown that users' primary motivation for use is convenience rather than security~\cite{amft2023would}, with some users even consciously avoiding storing credentials for high-value accounts in a password manager even as they use it for credentials for less sensitive accounts~\cite{amft2023would}. Users' thought process when choosing which password manager to use also tends to be driven by financial cost (e.g., if one requires a subscription fee) rather than security~\cite{munyendo2023just}. In practice, users frequently do not use password managers to their maximal security advantage and largely use them to autofill low-entropy passwords~\cite{amft2023would}. Moreover, users are generally still required to remember a password for the password manager itself.

\vspace{0.5em}
\noindent\textbf{Single Sign-On:}
Single-sign on (SSO) is a federated login technique that centralizes the responsibility for authenticating users with a single primary provider (most commonly Google or Apple~\cite{morkonda2021empirical}) using access delegation protocols such as OAuth and OpenID Connect.
SSO adoption has been limited by both legitimate privacy considerations over data sharing with big tech companies~\cite{sun2011makes,balash2022security,dimova2023everybody,morkonda2021empirical} and holdouts in adoption due to lack of trust in the underlying technology~\cite{sun2010billion,sun2013investigating}, with prior work showing users are less likely to use SSO for more sensitive accounts~\cite{cho2020will}. The crux of the privacy issue is that the centralized provider (e.g., Google) will be able to observe all authentication attempts for a particular user, though there have been several promising academic proposals to reduce data sharing~\cite{kalantari2023user,morkonda2022sign,guo2021uppresso,fett2015spresso}. Some platforms (such as GitHub) opted not to offer federated log-in to maintain greater control over the authentication process for their website~\cite{hill2017moving}.

There has also been a large body of academic work showing security vulnerabilities both in the underlying protocol~\cite{somorovsky2012breaking,sun2012systematically,wang2013explicating,mainka2017sok,mainka2017sok} and deployed implementations~\cite{wang2012signing,zhou2014ssoscan,yang2016model,wang2015vulnerability,mainka2016not,hu2014application,yang2018vetting,sun2012devil,bai2013authscan,ghasemisharif2022towards,ghasemisharif2018single,westers2023sso}, including real-world cybercrime networks that maintain honeypot websites and collect OAuth access tokens~\cite{farooqi2017measuring}.
Apart from specific security and privacy concerns, single sign-on schemes inherently present a single point of failure~\cite{herley2011research,bonneau2012quest} and hence an attractive target for attackers.

\subsubsection{Device-Bound Credentials}
\label{sec:device-bound-credentials}

With the increasing prevalence of hardware tokens and secure hardware chips in smartphones, authentication using only a single strong hardware-backed factor (the user's smartphone) is now viable---allowing providers to remove passwords from daily authentication flows. At a high level, device-bound credentials use public-key cryptography to authenticate, where the smartphone's secure hardware component generates a unique keypair for each web account, stores the private key in the hardware module, and shares the public key with the web server. To authenticate to the web service, a user need only authenticate to their local device authenticator using their regular device unlock mechanism (e.g., fingerprint, PIN, pattern); hence, this scheme is colloquially referred to as ``passwordless'' authentication. All major operating systems and web browsers now support passwordless authentication~\cite{fido_whitepaper22}, including cross-ecosystem authentication (e.g., Google to Apple) using either a QR code or Bluetooth Low Energy (BLE)~\cite{fido_whitepaper22,fido_cross_device} and cross-ecosystem sync using a platform-agnostic password manager such as 1Password~\cite{1password_passkey_sync}.

Device-bound authentication is made possible by the FIDO2 passwordless authentication protocol~\cite{fido2_protocol} and the WebAuthn standard~\cite{webauthn} for client-to-server communication, which were intended from the start as contenders for password replacement~\cite{fido_whitepaper22}. The initial deployed version of the Fast Identity Online (FIDO) protocol in 2019 (``FIDO U2F'') relied on two-factor authentication (2FA) hardware tokens (Section~\ref{sec:the-state-of-web-auth-and-recovery:primary-authenticaion:hardware-tokens}), and the passwordless standard was released in 2018~\cite{fido2_protocol}. In FIDO terminology, this means there are two ways for a user to authenticate: using a \textit{roaming authenticator}, such as a discrete hardware token, or a \textit{platform authenticator}, such as the built-in smartphone authenticator~\cite{fido_whitepaper22}.

\begin{figure}
    \centering
    \includegraphics[width=0.9\columnwidth]{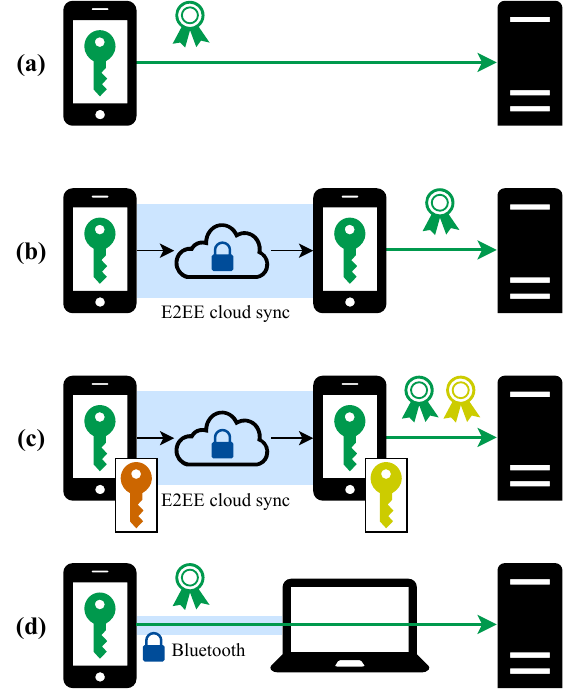}
    \caption{Passkey authentication flows: (a) standard authentication using a passkey on a mobile phone; (b) backup-eligible passkey (multi-device credential) is synchronized to a new device using an E2EE cloud backup service; (c) multi-device credential using the device-bound key extension to distinguish individual devices; (d) cross-device authentication on a laptop via Bluetooth.}
    \label{fig:passkey-examples}
\end{figure}

The primary benefit of device-bound credentials is mitigating phishing attacks and large-scale compromise since each account for each account has its own distinct keypair.
During login, the device signs a challenge to prove possession of the respective private key (see Figure~\ref{fig:passkey-examples}a).
Binding to the original web service ensures that authenticators do not sign challenges coming from malicious websites.
The primary downside is that credential security now reduces to device security~\cite{fido_whitepaper22}, as a compromised device can allow an adversary access to all services. Smartphone providers have deployed several security measures to prevent unauthorized access, such as requiring biometric authentication each time a passkey is used and instituting device unlock rate-limiting, but the precise security protections will depend on platform and user configurations.

Given that FIDO2 credentials were initially conceived of as being exclusively device-bound, recovery from device loss is a major concern for users even with smartphone-based authentication~\cite{wursching2023fido2,owens2021user}. While passkeys can be synced across devices, if all device(s) are lost, all passkeys are lost. When FIDO2 relied primarily on roaming authenticators in 2019, the FIDO Alliance's official recommendation was to register multiple authenticators for each account to prevent account lockout~\cite{fido_recovery_whitepaper19}, which was not viable for the general public. This continues to raise usability issues with smartphones as devices are frequently lost, stolen, broken, or wiped when upgrading to a new device. This is a common scenario: users may only own one device (or one FIDO-compatible device), a user may be traveling and only brought one device with them (e.g., left laptop at home and lost their phone). Industry consensus to address the recoverability problem is the \textit{passkey}.


\hfill \break
\noindent\textbf{Passkeys:}
Major industry vendors (namely, Apple and Google) refer to the FIDO2 passwordless authentication mechanism as \textit{passkeys}.~\cite{fido_whitepaper22} Google first introduced passkey support for Google Accounts in May 2023~\cite{brand_blog23} and Apple followed suit in June~\cite{davis_verge23}, though both will continue to support passwords as an authentication mechanism for the foreseeable future.

To enhance usability and mitigate passkey loss, major industry OS providers have implemented passwordless authentication to automatically sync credentials to the respective cloud backup service~\cite{google_blog_password_manager_backup,apple_keychain_recovery} (see Figure~\ref{fig:passkey-examples}b).
At a high level, cloud backup recovery works similarly for both Apple and Google: a user must first authenticate to the cloud service (e.g., Google Account or iCloud), and then further authenticate to the credential backup by providing their smartphone device unlock code~\cite{google_blog_password_manager_backup,apple_keychain_recovery}. While credential cloud sync is critical for satisfying end-user expectations of recoverability, cloud sync means that the practical security of passwordless credentials reduces to the security of the cloud account, including cloud HSM security~\cite{fido_whitepaper22,connell2024secret}, as is the case with traditional password managers~\cite{li2014emperor,silver2014password}.

\vspace{0.5em}
\noindent\textit{Apple:} On Apple devices, passkeys are part of the keychain information that can be backed-up and synchronized using iCloud~\cite{apple_passkeys}.
Passkey syncing to iCloud Keychain occurs by default but users can disable this synchronization for individual devices.
Keychain information is always backed-up using E2EE, where the decryption keys are stored in rate-limited hardware security modules (HSMs). To access the iCloud Keychain after authenticating to the iCloud account, Apple requires the user to enter either the device passcode, an alternately six-digit passcode previously selected by the user, or a longer auto-generated pseudorandom code, and limits the user to 10 attempts~\cite{apple_keychain_recovery}.

\vspace{0.5em}
\noindent\textit{Android:} On Android devices, passkeys are managed by the Google Password Manager which provides E2EE backups that can be restored on new devices.
To access backups on a new device, after authenticating to the primary Google account the user needs to enter the lockscreen pattern or PIN of an existing device.
Similarly to Apple, the backend uses HSMs to rate-limit attempts and to ensure that no information about the lockscreen authentication or the stored credentials is leaked.~\cite{google_blog_password_manager_backup}
When setting up a new device, users can transfer data locally from an old device which includes the stored passkeys.

\vspace{0.5em}
In addition to the OS passkey authentication process, the FIDO2 standard allows a domain to implement additional authentication steps after a user has authenticated using FIDO2 passkeys. For instance, the web service and and client can use an optional WebAuthn extension~\cite[\S10.2.2]{webauthn} to include an additional \textit{device-bound key} in the authentication process (see Figure~\ref{fig:passkey-examples}c), allowing the web service to, e.g., detect whether the user logs in from a new device~\cite{fido_whitepaper22}.
FIDO2 also allows for \textit{cross-device authentication} where authentication attempts on a new device, e.g. a laptop without passkey support, is tunneled to a device possessing a valid passkey for the service via Bluetooth (see Figure~\ref{fig:passkey-examples}d) using the Client to Authenticator Protocol (CTAP)~\cite{fido_ctap_21}.

Academic studies of end-user perceptions of passwordless authentication revealed inaccurate mental models (e.g., believing the fingerprint or other biometric data is sent to the web server~\cite{lassak2021s}) and challenges with inconsistencies in user interface design~\cite{oogami2020observation,owens2021user}, though passwordless authentication was generally considered simple to set up and use~\cite{wursching2023fido2}. Users repeatedly express concerns over account loss~\cite{wursching2023fido2,lyastani2020fido2}, though cloud sync recovery was not an option at the time the studies were conducted (and so users were presented with a design in which total device loss meant total account loss).

%

After reviewing usability studies of passwordless authentication and current FIDO2 deployments, we identify three residual concerns hampering passkey adoption:
\begin{enumerate}
    \item \textit{Credential Sharing:} Credential sharing is a common practice across numerous scenarios, including home and work envirnments~\cite{wursching2023fido2,wang2022s,lin2021s,kaye2011self,song2019normal,wu2022sok}. The latest industry deployments have made passkey sharing simpler with cross-ecosystem syncing, but there are still several cases where existing syncing mechanisms are inadequate~\cite{bicakci2022fido2}. One of the contexts in which credential sharing is most prevalent, the workplace, is particularly challenging given the recent trend towards remote work such that users may not be in physical proximity to each other, making passwords far simpler to share over common written communication channels.
    \item \textit{Credential Revocation:} The FIDO2 standard does not adequately address credential revocation at a global scale. Currently, it assumes the web services will implement the ability for a user to revoke access (e.g., remove the public key from the server) for specific authenticators, which a user will need to do for each individual website for which they have registered an authenticator~\cite{wursching2023fido2,lyastani2020fido2}. Revocation is a critical ability in several situations, including enterprise use cases with employee turnover, or intimate partner violence where the victim has previously shared a passkey with another device (e.g., their partner's device but now desires to revoke access), or when a device has been lost. A small academic work has begun to consider the challenge of global key revocation, in which a user desires to revoke all keys associated with a specific authenticator~\cite{hanzlik2023token}, but the current standard offers no satisfactory solution.
    \item \textit{Credential Availability:} That passkeys are bound to the trusted hardware of specific devices is a major advantage from a security standpoint but can pose a significant disadvantage for widespread usability and availability, particularly for at-risk demographics. Shared devices (including public computers~\cite{munyendo2023eighty}) are common, with users sharing devices for financial, cultural, and personal reasons~\cite{blumenstock2010mobile,karlson2009can,komen2016here,ahmed2019everyone,burrell2010evaluating,paudel2023deep}, including temporary sharing (e.g., showing a friend or relative a photo slideshow~\cite{karlson2009can,matthews2016she}). Conversely, this also raises privacy concerns over inadvertent account sharing depending on passkey access duration before requiring reauthentication in a particular implementation.
\end{enumerate}

\subsubsection{Social Authentication}
Social authentication and recovery, where an account holder designates one or more recovery contacts or ``trustees'' who can help them regain access, is the only recovery scenario that can dependably handle various real-world disaster cases (where the user loses or forgets all devices and passwords). First proposed by Brainard et al.~\cite{brainard2006fourth} of RSA in 2006, social recovery takes advantage of real-world trust relationships to authenticate a user by having a different user vouch for them. Brainard et al. motivated this concept of vouching by considering the financial cost to the company of password-reset staff, but in today's world social recovery is an important option because of the difficulty of authenticating the correct user as part of a manual recovery process, and because in some cases the provider may be unable to reauthorize the user (as in E2EE services), and has received renewed attention in recent cryptographic proposals~\cite{chase2022acsesor}.


\vspace{1em}
\noindent\textbf{Single Recovery Contact:}
The simplest social recovery scheme is to designate a single recovery contact (presumably a close acquaintance). Apple’s cloud backup scheme, introduced in 2022, deploys the idea of a recovery contact, described by Apple as ``a trusted friend or family member’’~\cite{apple_recovery_contact}. In Apple’s scheme, this contact is another iCloud user who generates a short code to send to an original user Alice enabling her to recover her account. In theory, a user with two iCloud accounts could also use one as the recovery contact for the other, though this arrangement would not adequately address the root concern. This is similar to a multi-wallet system in the cryptocurrency ecosystem, where a user leverages a secondary wallet to authenticate to the primary wallet \cite{rezaeighaleh2019new}. Service providers can also allow a user to designate multiple recovery contacts, where each contact has the ability to single-handedly provide the account holder with a code to restore access.

While Apple’s documentation promises that a recovery contact ``won’t have any ability to access your account, only the ability to give you a code to help you recover your account'', in practice a recovery contact could simply pretend to be the original user (assuming the contact knows a few additional basic facts, such as the iCloud account email used).

Apple takes several precautions to prevent a recovery contact from gaining access to the account. Apple requires any individual entering the recovery code to answer an additional set of security questions in the first instance. Most importantly, Apple documentation suggests there is a time delay between the recovery request and regaining account access, and specifies that users should not use their devices in the intervening period because to do so would indicate that the request is not legitimate. In short, a user who is logged in and actively checking their devices and/or accounts will be able to detect that such a recovery process has been initiated. If the user has associated other traditional two-factor authentication mechanisms with the account, such as a phone number or email address with a different provider, they may also receive a notification on this platform informing them of the access request.

\vspace{1em}
\noindent\textbf{Threshold Social Recovery:}
As an alternative to designating individual users as an all-powerful recovery contact, service providers can also offer a threshold secret sharing scheme where the secret is divided among multiple recovery contacts but can be reassembled once a certain number of shares are combined~\cite{shamir1979share}. In Shamir secret sharing, for instance, a key is divided into n pieces with a recovery threshold k such that k pieces (where k $\le$ n) are needed to reassemble a valid secret. This has the benefit of making the recovery mechanism more resilient against unavailable or malicious recovery contacts by distributing trust among multiple contacts and providing redundancy in social recovery.

Schechter et al.~\cite{schechter2009s} first proposed an account recovery scheme with multiple recovery contacts in 2009, where a key is split among the multiple contacts such that a minimum threshold of the total must share an account recovery code with the original user for them to regain access. Although Schechter et al. do not use the term Shamir secret and do not specify how the secret is divided up among the trustees, the scheme described functions in a similar manner to a standard cryptographic secret sharing scheme. Follow-on work has shown threshold trusted contact schemes are highly usable, with failures of trusted contact authentication due to time delays or timeouts (likely due to the perceived low value of the accounts used in experiments), rather than poor mental models or misconceptions~\cite{lassak2024comparative,schechter2009s,stavova2016codes}.

Secret-sharing schemes are already used in industry for recovering end-to-end encrypted data. PreVeil, a cross-platform cloud service offering end-to-end encrypted email and file storage for enterprises, deployed an opt-in threshold scheme in 2019, stating that a system that distributes trust among a set of trustees is more secure than one which centralizes trust in a single trustee~\cite{preveil_whitepaper}. In PreVeil’s design, when an account holder has initiated the recovery process, they will be presented with a list of their previously designated recovery contacts and able to select which members of the list should be used to approve the request. PreVeil’s system architecture is somewhat unusual in that it does not use passwords or require the user to enter any credentials in order to log in. Instead, they store the user’s private key on-device, allowing anyone authenticated to the device to access PreVeil storage. As a result, PreVeil needed a sufficiently failsafe backup mechanism in case the user loses their device(s) since no password or recovery code exists. Facebook used to offer a “Trusted Contacts” feature for regaining access to Facebook accounts (albeit not in an E2EE scenario) via a three-of-five secret sharing scheme~\cite{fb_trustedcontacts2023}, but deprecated the feature in 2022.

Social recovery has been gaining favor in the cryptocurrency ecosystem as well. In 2022 BitKey, a non-custodial hardware wallet (i.e., users store their own private keys), enabled an opt-in threshold social recovery scheme ~\cite{bitkey_recovery2022}, where an account holder designates three recovery contacts and two of the three are needed to restore access. PreVeil does not require a certain threshold size, but similarly specifies in their documentation that two-of-three is a typical setup.

\vspace{1em}
\noindent\textbf{Limitations of Social Recovery}
Social recovery goes a long way towards mitigating the challenge of an individual user managing their own keys, but at the same time presents several new concerns. While Apple takes several sensible precautions to prevent a recovery contact from gaining illicit access (instituting a time delay, requiring the individual requesting access to answer a series of additional security questions, etc.), this mode of recovery is nonetheless vulnerable in certain scenarios. We can reasonably assume that someone close enough to the account holder to be designated a recovery contact will likely be able to answer any additional verification information (including the email address associated with the iCloud account, date of birth, etc.), and therefore gain access to the account. An example would be an honest-but-curious recovery contact, such as a relative who initially requests access to recover family photos but later realizes that the account also contains years of messaging history.

A time-delay between the access request and when access is granted, during which the account holder is notified that a request has occurred, is essential for mitigating illegitimate access. However, time-delay schemes rely heavily on users’ attentiveness and assume that users check an account regularly, and recent academic work found that users often fail to recognize and respond to login attempt notifications~\cite{markert2024understanding}. Critically, the dependence on this proactive detection on the part of the user means that security guarantees of social recovery do not hold if the account owner has passed away. This is not a scenario most users contemplate for obvious reasons, but posthumous account access is fairly simple under Apple’s individual recovery contact scheme.~\footnote{Apple has a separate notion of a ``Legacy Contact'', an optional setting where a user’s legacy contact can recover account access by manually presenting a death certificate to Apple—but social recovery can intentionally or unintentionally also become a legacy contact.} In the case of a single recovery contact we must also consider an honest-but-curious recovery contact, such as a relative who initially requests access to recover family photos but later realizes the account also contains years of messaging history. A threshold secret sharing scheme would partly mitigate this scenario since multiple contacts would need to agree that access is acceptable.

Social recovery has also been exploited by online scammers. The process of receiving an unsolicited message from an acquaintance asking the recipient to click on a link and provide some information closely resembles real-world scams, an unfortunate reality which malicious actors used to their advantage. Facebook’s Trusted Contacts feature was the target of a popular scam in 2017 in which an attacker who has compromised a given Facebook account sent messages to the account owner’s contacts, pretending to be the owner and asking the recipient to click on a link to help them reset their password by providing the message sender with a recovery code~\cite{fb_contacts_phishing2017,javed2014secure}. Unfortunately for the victim, the link provided was in fact a password reset link for the recipient’s account, and the code the recipient sent back to the attacker allowed the attacker to compromise the recipient’s account as well. This scam possibly contributed to Facebook’s decision to no longer support trusted contacts as a recovery mechanism, though the company never officially provided a reason.

A more contemporary concern is that social verification may be vulnerable to manipulation by generative AI tools, such as a falsified video call or voicemail, with even close contacts unable to distinguish between genuine and artificial content. Simply speaking to another user over the phone was considered sufficient identification as part of a social recovery scheme as recently as 2016~\cite{stavova2016codes}, but there have been numerous voice-cloning attacks in recent years used in real-world scams~\cite{bethea_voice24,kassis2023breaking,abdullah2021sok}. Social authentication is all too easily susceptible to various social engineering attacks, such as where a contact calls from an unusual phone number and claims they have lost their smartphone and need assistance recovering an account---when in reality, the contact's voice is AI-generated.

\subsubsection{Long-Term Recovery Key}
There are several different terms for this concept (``recovery key'', ``recovery code'', ``master passphrase'', etc.), but they all refer to a pseudorandom string that the user presents to the service to restore access to their account. This recovery code is generally distinct from a standard user-generated password or PIN regularly used to log in in that it is arbitrary and usually substantially longer to guard against brute-force attacks.

The popularity of recovery codes as a recovery mechanism endures because they are generally the most secure and efficient way of recovering account access---provided a user stores the code in a safe place and doesn't lose it. Virtually every cloud backup option either requires users to use a recovery key of some sort or offers it as an option if their protocol allows for multiple recovery mechanisms: Apple iCloud lets users use a 28-character recovery key (in addition to the ordinary iCloud password)~\cite{apple_recovery_key}, and WhatsApp encrypts backups with either a 64-digit encryption key or a user-generated password~\cite{whatsapp_e2ee_backup}. Meta's Labyrinth protocol offers several different recovery mechanisms, one of which is a standard 40-character recovery code~\cite{messenger_whitepaper}. Signal does not support cloud backup but lets users encrypt a local backup using a 30-digit recovery key which the user is responsible for storing and safeguarding. These codes are generally only shown once upon creation, although a logged-in user can usually also generate a replacement recovery code even if they have lost the old one.

A recovery key option suffers from the same problem as a user-generated password: users all too often forget or lose it. A small number of users may neglect to save it at all, often out of overconfidence that they will never need it~\cite{holtervennhoff2024mixed}. Unintentional loss is even more likely given that the nature of a recovery key is that it is used rarely, if ever. A user may store it on local device storage or on a physical piece of paper, encounter it many months later, and throw it out without realizing its significance. If the user's only backup recovery mechanism is this passcode (in addition to losing access to their device and/or regular password), they have no recourse and are locked out of their account permanently. The simplest mitigation technique is to create redundant copies of the passcode, though this increases the attack surface and potentially requires users to store the passcode where family members or others can access it.

\subsubsection{Manual Recovery}
Manual recovery (``ad-hoc schemes''~\cite{holtervennhoff2024mixed}) are the recovery scheme of last resort~\cite{lassak2024comparative,parkin2016assessing,foo2000integrated,gerlitz2023adventures}. Some large industry providers offer a formally described manual recovery process~\cite{google_manual,apple_manual}, while most others offer generic support contact information. Providers may also offer appeals processes in cases where a provider's content moderation scheme flags the account~\cite{nyt_csam2022}. On a large scale, however, there is little incentive for small service providers to expend significant effort of these types of schemes, especially for users of unpaid services. Importantly, provider-assisted recovery is inherently not possible in E2EE services, which usability research has shown that some users do not understand, with users of an E2EE email service mentioning provider assistance as a possible recourse after recovery code loss~\cite{holtervennhoff2024mixed}.



\vspace{.5em}
\noindent\textbf{Break-Glass Encryption} To develop an E2EE variant of manual recovery, a recent thread of academic work has attempted to tackle challenges around encrypted data loss by focusing on detecting, rather than outright preventing, account access~\cite{scafuro2019break}. Orsini et al.~\cite{orsini2023recover} proposed a cryptographic scheme for emergency access to cloud data storage, using the same threat model as in this work where a user Alice has lost all relevant credentials and all devices. They propose a credential-less authentication scheme in which any user can request access to a cloud account knowing only the associated email address or similar username, but there are only two possible states for a given account: either the legitimate user Alice is logged in and can monitor and reject illegitimate access requests within a certain timeframe, or Alice has become locked out of her account (e.g., by losing her device) and her request to regain access will be automatically granted after some time period has elapsed (since there is no legitimate user to reject it).

Such schemes are entirely dependent on detectability: the assumption is that the legitimate user will be consistently online, and confidentiality is guaranteed by proactive action on the part of the account owner. Both the Orsini et al. scheme and a similar concept for cryptocurrency wallets~\cite{blackshear2021reactive} assume an information asymmetry between the legitimate user and all other users in that the legitimate user would know when they have lost access (and request to be restored to the account) before anyone else, but this does not always hold (e.g., a device is stolen, posthumous access, etc.). Perhaps the biggest concern with these ``break-glass encryption'' schemes is that a deceased or incapacitated account holder is now vulnerable to any relatives or acquaintances familiar with their account name to a far greater extent than was the case with existing social recovery schemes. We are skeptical that any such scheme would ever be feasible for the general public.

\subsection{Secondary Authentication Mechanisms}
\label{sec:secondary-auth-mechanisms}
Common second factor authentication (2FA) mechanisms used historically and still today include recovery questions, email and SMS-based 2FA, authenticator apps, hardware tokens, and risk-based authentication (e.g., using browser metadata to flag suspicious login attempts). As these techniques have been covered extensively in prior work, we defer an in-depth discussion of them to Appendix~\ref{sec:secondary-auth}.

\section{Passkey Deployment}
Here, we investigate the deployment of passwordless authentication in practice since the FIDO2 standard was published in 2019~\cite{fido2_protocol}.

\subsection{Methodology}
To measure passkey deployment, we manually inspect each site in the top 200 of the Tranco dataset of most widely visited domains~\cite{pochat2018tranco}. For sites that offer multiple account versions (e.g. a free version and a paid version), we follow the methodology of Gavazzi et al.~\cite{gavazzi2023study} and select what we expect to be the most common account type. All experiments were conducted by logging in from Chrome 124.0.6367.203 on macOS Sonoma 14.2.1.

\subsection{Passkey Availability}
Of the top 200 domains, we successfully audited $n = 94$ (the remaining sites either did not offer account creation, required service-specific information to set up the account such as a phone number with a particular provider, or did not load). Of these 94 sites, we find 28 sites (29.8\%) support passkeys for user authentication. We observe that a further 31 sites (33.0\%) do not directly support passkeys but offer single-sign on with a provider which \textit{does} support passkeys (in the vast majority of cases, Google). If we include sites which offer indirect passkey support through SSO, 59 of 94, or 62.8\%, of sites in the top 200 directly or indirectly offer passkey support, a significant increase from a 2021 study which found no support for passwordless authentication among 235 popular sites~\cite{gavazzi2023study}.

\section{Keys Not Under Doormats: Recovery in End-to-End Encrypted Systems}
End-to-end encryption (E2EE) is increasingly being deployed to secure long-term data storage through cloud-based services as well as communications. Traditionally mentioned in the context of secure messaging, E2EE can be deployed for many scenarios where a user wants to prevent their data being visible to a third party (namely, the service provider storing the data, though the provider will still retain user metadata). But E2EE comes at a price: if only the user has all the information needed to access the data, the service provider is unable to come to the rescue should a user forget their password or lose their client device. With E2EE messaging and email services, loss of conversation history, while irritating, has generally been an acceptable consequence if a user sets up a new phone and has forgotten the password to their E2EE backup.

The prospect of E2EE for cloud backup services such as Apple iCloud make security and recoverability trade-offs more salient. Users store troves of important photos, documents, and other valuable long-term data in cloud services.

\begin{table*}[]

\begin{tabular}{@{}lcccccccccc@{}}
                       & \multicolumn{1}{l}{}              & \multicolumn{9}{c}{\textbf{Authentication and Recovery Mechanism}}                                                                                                                                                                                                                                                                                                           \\ \midrule
                       & \multicolumn{1}{l|}{\textbf{E2EE Platform}}     & \begin{tabular}[c]{@{}c@{}}Device\\Keychain\end{tabular} & \begin{tabular}[c]{@{}c@{}}User-Chosen\\Password\end{tabular} & \begin{tabular}[c]{@{}c@{}}Recovery\\Code\end{tabular} & \begin{tabular}[c]{@{}c@{}}Third-Party\\Storage\end{tabular} & \multicolumn{1}{l}{PIN}   & \begin{tabular}[c]{@{}c@{}}Recovery\\File\end{tabular} & \begin{tabular}[c]{@{}c@{}}Recovery\\Email\end{tabular} & \begin{tabular}[c]{@{}c@{}}Recovery\\Contact\end{tabular} & \begin{tabular}[c]{@{}c@{}}Recovery\\Group\end{tabular} \\ \midrule
                       & \multicolumn{1}{c|}{Apple iCloud} & \fullcirc            & \fullcirc                    & \fullcirc          & \emptycirc               & \emptycirc & \emptycirc         & \emptycirc          & \fullcirc             & \emptycirc          \\
                       & \multicolumn{1}{c|}{NordLocker}   & \emptycirc           & \fullcirc                    & \fullcirc          & \emptycirc               & \emptycirc & \emptycirc         & \emptycirc          & \emptycirc            & \emptycirc          \\
                       & \multicolumn{1}{c|}{pCloud}       & \emptycirc           & \fullcirc                    & \fullcirc          & \emptycirc               & \emptycirc & \emptycirc         & \emptycirc          & \emptycirc            & \emptycirc          \\
\textbf{\begin{tabular}[c]{@{}l@{}}Storage\end{tabular}}      & \multicolumn{1}{c|}{MEGA}         & \emptycirc           & \fullcirc                    & \fullcirc          & \emptycirc               & \emptycirc & \emptycirc         & \emptycirc          & \emptycirc            & \emptycirc          \\
                       & \multicolumn{1}{c|}{Tresorit}     & \emptycirc           & \fullcirc                    & \emptycirc         & \emptycirc               & \emptycirc & \emptycirc         & \emptycirc          & \emptycirc            & \emptycirc          \\
                       & \multicolumn{1}{c|}{Internxt}     & \emptycirc           & \fullcirc                    & \fullcirc          & \emptycirc               & \emptycirc & \emptycirc         & \emptycirc          & \emptycirc            & \emptycirc          \\
                       & \multicolumn{1}{c|}{Filen}        & \emptycirc           & \fullcirc                    & \fullcirc          & \emptycirc               & \emptycirc & \emptycirc         & \emptycirc          & \emptycirc            & \emptycirc          \\ \midrule
                       & \multicolumn{1}{c|}{Proton}       & \fullcirc            & \fullcirc                    & \fullcirc          & \emptycirc               & \emptycirc & \fullcirc          & \emptycirc          & \emptycirc            & \emptycirc          \\
\textbf{Email} & \multicolumn{1}{c|}{PreVeil}      & \fullcirc            & \emptycirc                   & \emptycirc         & \emptycirc               & \emptycirc & \fullcirc          & \emptycirc          & \emptycirc            & \fullcirc           \\
                       & \multicolumn{1}{c|}{Tutanota}     & \emptycirc           & \fullcirc                    & \fullcirc          & \emptycirc               & \emptycirc & \emptycirc         & \emptycirc          & \emptycirc            & \emptycirc          \\
                       & \multicolumn{1}{c|}{StartMail}    & \emptycirc           & \fullcirc                    & \fullcirc          & \emptycirc               & \emptycirc & \emptycirc         & \fullcirc          & \emptycirc            & \emptycirc          \\ \midrule
                       & \multicolumn{1}{c|}{FB Messenger} & \fullcirc            & \fullcirc                    & \fullcirc          & \fullcirc                & \fullcirc  & \emptycirc         & \emptycirc          & \emptycirc            & \emptycirc          \\
\textbf{Messaging}              & \multicolumn{1}{c|}{WhatsApp}     & \fullcirc            & \fullcirc                    & \fullcirc          & \fullcirc                & \emptycirc & \emptycirc         & \emptycirc          & \emptycirc            & \emptycirc          \\
                       & \multicolumn{1}{c|}{Signal}       & \fullcirc            & \emptycirc                   & \emptycirc         & \emptycirc               & \emptycirc & \emptycirc         & \emptycirc          & \emptycirc            & \emptycirc          \\ \bottomrule
\end{tabular}

\vspace{1em}
\caption{\label{tbl:e2ee_mechanisms} Deployed recovery mechanisms for end-to-end encrypted storage, email, and messaging services. The table displays all recovery options offered by each service, but in practice some of these choices may be mutually exclusive. For instance, WhatsApp allows a user to set either a recovery code or a user-chosen recovery password, but not both.}

\end{table*}

\subsection{Move Towards E2EE Storage}
Historically, cloud service providers have opted not to encrypt account data storage end-to-end due to political pressure~\cite{abelson2015keys,abelson2021bugs} and concerns over usability and account lock-out. Academic studies on the usability of multi-factor authentication apps have found account lockout to be a repeated concern for users since app backups are generally encrypted~\cite{gilsenan2023security,reynolds2018tale}. The risk users may lose long-term data is likely part of the reason that E2EE storage has been slow to catch on even among providers who deployed E2EE for messaging several years prior. WhatsApp, for instance, rolled out E2EE communications in 2016, but did not offer E2EE backups of these communications until 2021~\cite{whatsapp_e2ee_backup2}.

In the wake of a changing technical and political landscape, however, Apple has publicly advocated for protecting cloud data with end-to-end encryption~\cite{krstic_icloud2023} and rolled out an opt-in E2EE backup scheme in late 2022. Their public justification for making E2EE backup opt-in for users, rather than the default, is to reduce the risk of permanent data loss as ``the feature requires the user to take ultimate responsibility for managing their cryptographic keys''~\cite{krstic_icloud2023}. While Apple's deployment represents a major step forward for cloud data privacy, encrypted data storage for the general public requires us to revisit what security-usability trade-offs are acceptable since a provider by definition cannot restore access. A well-designed user interface can warn the user of the consequences if they lose their recovery credentials, but this could also serve to deter users from opting in. To encourage widespread adoption of E2EE storage, we need schemes which are suitable for the average user.

Encrypted data storage is similar to non-custodial wallets in the cryptocurrency ecosystem, a space notorious for tales of irreversible data loss. Non-custodial wallets require the user to control their own private keys, instead of having the keys managed by their service provider~\cite{erinle2023sok}. The corollary to this setup is that the user has no recourse if they lose their private key, leading to permanent loss of the currency stored in the wallet. Cryptocurrency firms have been grappling with this problem for years, though the problem setup is slightly different in that cryptocurrency storage is often accessed infrequently, while messaging and storage applications can be accessed multiple times per day (which can affect which schemes are considered viable for a general userbase).


\subsection{E2EE Storage Deployments}
Apple iCloud, Facebook Messenger, and WhatsApp have all deployed opt-in end-to-end encrypted backup services in the past two years. In 2022, Apple's Advanced Data Protection scheme provides users with the option to encrypt their iCloud backups end-to-end, such that the encryption keys are replicated across all user devices. Previously, it had only been possible to encrypt most iCloud data such that Apple retained access. 
While several other services already offered E2EE storage, these services are largely targeted at a more tech-savvy userbase and have not seen mass adoption.

Specific protocols vary, but providers generally achieve E2EE backup by storing decryption keys in hardware security modules (HSMs) such that the user authenticates to the HSM and the provider relays messages between the client device and HSM but has no access itself. While this represents a major improvement in the security level offered by these services, this development raises natural follow-up questions around recovery and usability.

Facebook Messenger's E2EE protocol, Labyrinth~\cite{labyrinth_whitepaper}), represents an interesting case study of trade-offs between authentication and data recovery. 
Since Messenger is commonly used as a web application, Labyrinth is designed to allow users to log in on a new device or browser using only their ordinary Facebook credentials. If a user no longer has access to their cryptographic key material (namely, their private authentication key), they will still be able to log in to their account, but will not have access to their conversation history. This is a reasonable trade-off for E2EE messaging, but does not work for E2EE of long-term storage, where the whole purpose is to access previously generated data.

\subsection{E2EE Storage Recovery Mechanisms}
Here, we outline and discuss deployed recovery mechanisms in E2EE schemes, systematizing currently deployed protocols across end-to-end encrypted storage, messaging, and cryptocurency wallets. Table~\ref{tbl:e2ee_mechanisms} shows the diverse array of recovery mechanisms used by the 14 most widely used E2EE providers for storage, email and messaging. Our discussion focuses on services designed for individual consumer use rather than services targeted at enterprises.

If a user has access to a logged-in client device, recovery is simple. A common recovery mechanism invisible to the user is to automatically save a decryption key to the browser or device’s local keychain where it can be accessed upon device unlock, enabling the user's device to serve as an authentication mechanism. WhatsApp, for instance, allows a user to reset their recovery code through the WhatsApp app after authenticating to their device using biometrics or entering the device PIN (which allow the WhatsApp client to access the encryption stored on device) \cite{whatsapp_pw_faq}. Apple's Advanced Data Protection E2EE cloud backup scheme offers a similar recovery protocol \cite{apple_pw_faq}, and Messenger's Labyrinth protocol allows users to send a one-time code to their old device~\cite{labyrinth_whitepaper}.

The challenging scenario is the case where a user has lost access to both the password used to unlock the account in question and, where applicable, all relevant client devices. For instance, perhaps someone has lost their phone, and attempts to restore WhatsApp on a new phone, only to discover they cannot find the decryption key to the WhatsApp backup.

We find that recovery codes are the primary backup method used to recover access to encrypted data, with 11 of the 14 providers surveyed offering or mandating this backup method. Unfortunately, it is all too easy for users to make mistakes that can lead to accidental account lockout, such as taking a screenshot of the recovery code which is then synced to the cloud storage to which the code restores access~\cite{holtervennhoff2024mixed}. An additional consideration is that non-enterprise cloud services are often used by small businesses and community organizations in addition to personal usage, and access credentials may have high turnover in ways that can make it easier for legitimate users to forget or lose access to credentials.

Recent academic research has suggested that recovery code loss is not uncommon in practice: Holtervennhoff et al.~\cite{holtervennhoff2024mixed} investigated how users perceive and store recovery codes in E2EE services, conducting a user survey of 281 users of Tutanota, an E2EE email service, and qualitatively analyzing Reddit support threads for the same service. They found several support threads in which users had lost their account password, 2FA device, and recovery code, and the user survey revealed a small number of users reported that they had not recorded the recovery code at all, believing there was no chance they would lose their passwords. Approximately 12\% of users surveyed believed Tutanota could help them regain access in case of recovery code loss, and only 14.8\% of users saved the recovery code in more than one location. These results are derived from the userbase of a privacy-centered email service, which the authors acknowledge ``is not representative of email users or privacy-conscious users in general''~\cite{holtervennhoff2024mixed}, so we may anticipate user misconceptions to be even higher in a service targeted at a mass audience.

As a final form of fallback, we note that StartMail, an E2EE email service, offers users the ability to request an email reset~\cite{startmail_recovery}. To preserve E2EE, on the backend their scheme requires the keys of two separate staff members to jointly recover the user's lost decryption key. While this scheme is not properly E2EE in the sense that colluding employees can access user keys, it nonetheless presents an interesting case study.



\subsubsection{Human-memorable passcodes} To explore more usable solutions, some providers allow users to enter a shorter or more memorable recovery code, such as a seed phrase, a short PIN, or a user-chosen password. To avoid brute-force attacks, these low-entropy codes are then used to generate a longer pseudorandom string.

Meta's Labyrinth design, for instance, allows users to enter a 4-digit PIN and then uses this short PIN to authenticate to a pseudorandom recovery code stored in a rate-limited HSM, limiting the number of attempts to 10. Similarly, Dauterman et al.~\cite{dauterman2020safetypin} had previously proposed SafetyPin, a backup system for mobile devices that encrypts the backup using a short (four or six digit) device PIN, but introduces additional protection from brute-force attacks and HSM compromise. Even with rate-limiting, however, PINs may still be easily guessed as users are also liable to choosing easy-to-guess PINs or PINs based on easily discoverable dates such as a birthday~\cite{markert2020pin}. PINs are also potentially compromised via social attacks such as shoulder-surfing~\cite{wsj_iphone_theft2023} and thus rate limiting mitigates but does not prevent even external attackers from compromising a numeric PIN.

Short, numeric PINs are also not a foolproof strategy to counter the fallibility of human memory as some percentage of users will still invariably lose access to the PIN. In one study of Signal PINs, 12\% of participants reported ``occasionally, frequently, or very frequently'' forgetting their PIN~\cite{bailey2021have}.

WhatsApp's E2EE backup scheme offers the option of a user-generated password to authenticate to the pseudorandom recovery code instead of directly storing the recovery code, where the low-entropy password is used to generate the proper 64-digit key using an oblivious pseudorandom function. 
From the user's perspective, this may be simpler to use as they only need to enter the shorter or more human-memorable sequence. Further variations include a ``recovery phrase'', a long string of between 12 and 24 words that the user can either store or attempt to memorize (typically referred to as a ``brain wallet'' by cryptocurrency service providers~\cite{erinle2023sok}). Proton, an end-to-end encrypted email service, uses a 12-word recovery phrase instead of a more conventional pseudorandom recovery code~\cite{proton_recovery}. The length of recovery phrases, though, makes memorizing these phrases an unattractive option for the general public, with the net result that recovery phrases offer little to no benefit compared with a standard pseudorandom string.

\subsection{Takeaways from E2EE Recovery}
Having observed that recovery codes are the primary backup method used to recover access to encrypted data, we discuss potential design modifications to improve end-user recovery.

\subsubsection{Usability Improvements}
There are additional feature choices providers can offer to mitigate the risk that a user loses or forgets their recovery passcode in the first place. WhatsApp and Signal both provide regular password reminders asking users to confirm their backup PIN, though in Signal’s case a user optionally has to enter a short PIN while WhatsApp requires users to enter either their password or full 64-digit code before they can access the app~\cite{whatsapp_reminder}. Other providers may offer a similar feature as well, though it is not explicitly stated in the documentation. One study of Signal's opt-in PIN reminder, however, found that roughly quarter of participants in one survey reported that they rarely or never confirmed their PIN when prompted~\cite{bailey2021have}. While mandatory reminders may be more effective, they also run the risk of becoming a nuisance to users, potentially leading users to disable E2EE storage to avoid these notifications and achieving the opposite of the desired effect.

\vspace{0.5em}
\noindent\textit{Cross-Provider Syncing:} A promising feature from a usability standpoint is the ability to automatically store the recovery key in a separate cloud service. Meta’s Labyrinth protocol gives users the option to store their pseudorandom recovery code in either Google Drive or iCloud Drive (depending on the mobile platform), where it is stored in a hidden folder in the third-party cloud service and does not preclude the user from separately storing the recovery code elsewhere as well. This form of automated storage represents an improvement from a usability standpoint in that a user would now have to lose access to both cloud providers, but there are privacy considerations: in a study of 2FA backups, Gilsenan et al.~\cite{gilsenan2023security} found that automatically uploading backups to Google Drive requires the user to grant read access to the additional service for their Google account name, email address, and photo.

Proton provides a slight twist on automatic storage by offering a platform-agnostic encrypted ``recovery file'' that can be stored long-term and provided to Proton at a later date to restore access~\cite{proton_recovery}. The documentation rather cryptically suggests that both recovery phrases and/or files ``may become outdated'', at which point a user will be warned that this recovery mechanism is no longer valid to restore access but will have to generate a new recovery phrase or file (though it is unclear from the documentation why or how often this might occur). In general, automatic cross-provider cloud storage represents a real usability improvement over asking the user to store the key on separate physical hardware (e.g., a USB stick) or to write it down on a piece of paper somewhere, since both are easy to lose accidentally.

\subsubsection{Increased Recovery Options}
Above all, the expectation that users will handle retaining their own decryption keys, the predominant strategy still in use today, may be a fair assumption with the risk assessment of E2EE messaging backups, but does not pass muster with general cloud storage backups. Cloud service providers may be reluctant to offer certain recovery options out of concern over increasing the attack surface. But E2EE cloud storage with more usable recovery options is still more secure than non-E2EE cloud storage, where a user is \textit{guaranteed} that at least one third-party, the service provider itself, can access their account data. Perhaps most importantly, the variety and lack of consensus around features make it even more challenging for an end user to confidently opt-in, ultimately hindering widespread E2EE adoption.

\section{Literature Survey}

We compiled relevant literature on specific authentication schemes found in contemporary web authentication (including smartphone authentication) by searching a range of keywords (including authentication, password, 2FA, MFA, recovery, security key) across relevant academic venues. We search the four major security conferences (USENIX Security, IEEE S\&P, NDSS, CCS) as well as other highly relevant conferences and workshops (PETS, SOUPS, CHI) and collect all papers published since 2012 (a year chosen to reflect changes in available authentication techniques since the comparative evaluation framework introduced by Bonneau et al.~\cite{bonneau2012quest}). 

Since there are numerous potential spelling variations (e.g., ``authenticator'' instead of ``authentication'', or ``Multi-Factor Authentication'' instead of ``MFA''), to augment our keyword search results we additionally search through citations of a subset of seminal papers via Google Scholar, including highly-cited works published at a small number of additional venues, to ensure we capture the vast majority of relevant work. After manual inspection of titles and abstracts, we exclude papers on subjects adjacent but not directly related to one of the categories of end-user authentication schemes discussed in Section~\ref{sec:the-state-of-web-authentication-and-recovery} (e.g., phishing, client-to-server authentication protocols). We further generally exclude academic proposals of novel authentication schemes unless the work contains a usability study or broadly applicable lessons for deployed schemes.

\begin{figure}[]
\begin{center}
    \includegraphics[width=\columnwidth]{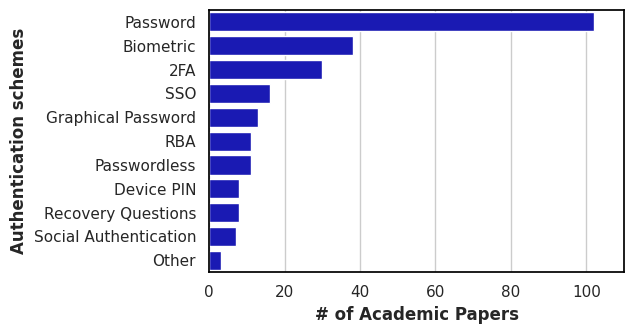}
\end{center}
\caption{Literature survey of academic work on real-world authentication schemes.}
\label{fig:lit_survey}
\end{figure}

\subsection{Results}
\label{sec:lit_survey_results}

Our literature search resulted in 245 papers that we broadly categorize by authentication scheme as shown in Figure~\ref{fig:lit_survey}. Table~\ref{tbl:lit_search} in the Appendix shows the specific papers included in each category. A small number ($n = 2$) papers were classified under more than one category, so the sum of the individual categories is slightly higher than the overall total.

Overall, we find that 41.6\% of academic authentication research has focused on passwords, including password usability, password managers, measuring password reuse, and various other aspects. Importantly, we find just 8 papers (3.3\%) focused on smartphone device PINs (a critical component of passwordless authentication recovery under Google and Apple's designs), only 7 papers studying social authentication in some capacity (a scheme deployed in widely used E2EE services), and just 1 paper (categorized under `Other') exploring the usability of E2EE recovery codes.

\section{Key Findings}
We identify seven key findings (KFs) from our systematic review of the security, usability, and privacy properties of contemporary authentication schemes.

\vspace{0.5em}
\noindent\textbf{KF1:} \textit{Increasing trend towards single-point-of-failure schemes.} From the perspective of daily usage, the most usable authentication schemes tend to be heavily centralized to minimize the amount a user has to memorize or carry with them (e.g., a single password manager protected by a master password, all passwordless authentication credentials backed up in a cloud service, SSO schemes that centralize sign-on with a single provider, 2FA apps that exist only on a user's smartphone). From a recovery perspective, these schemes represent a single point of failure. Even as users report centralized schemes to be highly usable on a day-to-day basis, they simultaneously cite fear of device loss as a persistent theme across studies of possession-based authentication mechanisms (e.g., smartphone containing 2FA app). Account loss with a cloud service provider can be devastating: stories of users who had their accounts shut down after being unfairly flagged by Google as uploading inappropriate material~\cite{nyt_csam2022,nyt_csam2023} vividly illustrate the practical consequences of sudden account loss, with one such user describing that it ``felt as if her house had burned down''~\cite{nyt_csam_appeals}.

From a security perspective, as the value of a particular account increases as credentials are centralized with a small number of providers (and possibly only a single provider), the effort attackers will be willing to devote to compromise the account will similarly increase. The net result of this centralization is that compromise of the master account will prompt cascading account compromise for all other credentials stored within the master account, particularly given that users frequently misconfigure relationships between accounts (e.g., chaining or looping account access) and underestimate the interdependency of their accounts~\cite{kraus2021users,hill2017moving,pohn2023framework,hammann2019user,hammann2022}.




\vspace{0.5em}
\noindent\textbf{KF2:} \textit{Device loss should be a basic assumption underpinning any threat model for device-bound credential authentication.}
Industry recovery schemes have gravitated towards device-based authentication as a user-friendly authentication technique where authentication keys are stored in the device keychain in E2EE recovery and elsewhere~\cite{google_passkey}. Such schemes further create a single point of failure as devices (particularly mobile devices) are frequently lost, changed, or stolen~\cite{gerlitz2023adventures,hang2015locked} and usability studies of smartphone authenticators have identified smartphone loss or unavailability as a frequent concern for users~\cite{owens2021user}, but threat models often assume multiple devices or simply don't account for benign (and frequent) threats such as accidental loss or damage~\cite{grosse2012authentication,grassi2020digital,frymann2020asynchronous,takakuwa2019moving}.

At minimum, usable recovery solutions must operate under the assumption that users may lose at least one (and possibly more) of their recovery devices. Given that the primary motivation for backing up in the first place is in case the device is lost~\cite{green_icloud2012}, a realistic recovery threat model should \textit{assume} that the physical device has been lost, rather than treating this as an exceptional case.

Redundancy across multiple devices (e.g. when a user is authenticated to an account on both a laptop and mobile device) is an important mitigation technique, but it is common for users to only access a service on a single device. For instance, E2EE messaging applications are frequently used only on a mobile device, and certain encrypted storage services are designed to be used exclusively as web applications. Many users do not own multiple devices, or multiple FIDO-compliant devices: 15\% of smartphone owners in the U.S. are ``smartphone-only'' Internet users and do not have broadband service at home~\cite{pew_mobile2024}. Even users who own multiple devices may go for extended periods of time in possession of only one when traveling (e.g., leaving the laptop at home and bringing only a mobile device while on vacation). Finally, real-world disaster scenarios that can cause total device loss (such as a house fire or natural disaster) seem implausible until they occur, compounding the loss of physical possessions with the loss of digital data~\cite{urbina_passwords2014}.



\vspace{0.5em}
\noindent\textbf{KF3:} \textit{``Passwordless'' authentication schemes invariably require a single master password.} The design paradigm of passwordless authentication recovery is to back up device-bound credentials to a cloud service (e.g., iCloud or Google Password Manager, depending on the device OS), encrypt the backup, and require the user to remember or store the cloud account password or device passcode. 

This design brings us to the same usability issues as with password managers: when all credentials are collected and stored in a single place with a master password, everything hinges on the user remembering this password. That the user will always need to provide \textit{something} to the provider is a wicked problem in authentication that comes up in all schemes~\cite{bicakci2022fido2}, but we know from usability research that users resort to insecure coping mechanisms, which in this case could be choosing a less secure password or passcode to ensure they don't forget it, limiting the benefits of rate-limiting guessing attempts. Additionally, passwords that are entered less frequently will likely be more difficult for users to memorize~\cite{bonneau2015passwords}, though providers may be able to mitigate this to some extent with regular re-entry reminders.


\vspace{0.5em}
\noindent\textbf{KF4:} \textit{E2EE contexts require a wider set of authentication schemes than general web authentication.} Given that E2EE inherently lacks provider-assisted recovery as a fallback option, we observe clear differences between authentication schemes deployed in E2EE compared with schemes deployed in general use cases. In particular, while none of the top-100 websites currently offer recovery via trusted contacts, multiple E2EE services offer some form of social authentication recovery (and in the case of PreVeil, trusted contact recovery is the \textit{only} option). Our survey of existing academic literature reveals very few papers investigating trusted contact recovery, despite the fact that this scheme is deployed in large-scale systems (and almost simultaneously become vulnerable to recent advances in generative AI). In this respect, academic focus has not kept pace with industry deployments.

We find that service providers too often lean towards prioritizing security over prioritizing usability, relying on a PGP-era approach of expecting users to copy and store a pseudorandom decryption key as a failsafe. There is little to no consensus around alternate recovery methods beyond the standard lengthy recovery key, with practically each major provider offering different features. Failure to provide users with a greater diversity of recovery options (which users have expressed a desire for~\cite{holtervennhoff2024mixed}) may make non-technical users reluctant to opt in to advanced security schemes like enabling E2EE backup out of fear of becoming permanently locked out of their digital house. The net result may be that E2EE is limited to a smaller subset of dedicated users, ultimately limiting the security benefits for the general public.


\vspace{0.5em}
\noindent\textbf{KF5:} \textit{Recovery schemes are perceived as exceptional scenarios by the end-user but as primary targets by attackers.} The importance and value generally increases during the account lifetime as more data is stored (and in a twisted corollary, the likelihood that passwords and recovery keys will be forgotten or lost increases proportionately). To reflect heightened data value (and how this value increases the longer an account has been in use), providers should offer users a wider selection of recovery options along the security-recovery spectrum. Users may also be more willing to accept additional risk-based authentication measures to reduce the likelihood that their account could be inadvertently compromised for high-value accounts. Service providers can use the comparative infrequency of recovery mechanisms and user's desire to regain access to their advantage by instituting additional measures, such as an access time-delay or account notification, to maintain a similar level of account security without sacrificing usability.



\vspace{0.5em}
\noindent\textbf{KF6:} \textit{Inconsistency in categories of provided recovery schemes across 2FA apps and E2EE systems harms both usability and security for end-users.} Prior work has identified inconsistencies in user interface design to be a potential factor limiting users' adoption of a particular scheme~\cite{lassak2024aren,lassak2021s,golla2021driving,ghorbani2023systematic}, including in passwordless authentication schemes~\cite{bicakci2022fido2}. But inconsistencies across which authentication and recovery possibilities are provided in the first place is equally as important for accurate user mental models and understanding of recovery possibilities. Inconsistencies in 2FA backup recovery processes make it more likely a user may misconfigure backups~\cite{gerlitz2023adventures,amft2023we}, and recent work showed that some users believe common provider-assisted recovery (e.g., password reset, SMS OTP) is possible in E2EE~\cite{holtervennhoff2024mixed}

To enable passkey adoption on a large scale, there is a need for industry to standardize available recovery options across E2EE credential storage and web services, particularly for providers that serve a user base comprised of the general public (e.g., Apple). At the same time, services should offer a variety of recovery methods to cover the diversity of users’ personal situations, and offer the ability to enable more than one recovery path for sufficient redundancy. We find that particularly in E2EE services providers are still offering a comparatively small selection of reset options, particularly considering the relative value of some accounts to the user. ~\cite{lassak2024comparative} also called for multiple reset options.

\vspace{0.5em}
\noindent\textbf{KF7:} \textit{Contemporary recovery schemes generally do not reflect patterns of user behavior observed repeatedly in the academic literature.} Users have mentioned credential sharing as a common and desirable property in both personal and work contexts in usability studies of various authentication schemes, including passwords~\cite{adams1999users,inglesant2010true,singh2007password,theofanos2021passwords,urbina_passwords2014}, security tokens~\cite{reynolds2018tale}, 2FA apps~\cite{colnago2018s}, and passwordless authentication~\cite{kepkowski2023challenges}. Enterprises have been reluctant to deploy passwordless authentication as the FIDO2 protocol lacks native support for several common enterprise scenarios, including credential delegation and sharing~\cite{kepkowski2023challenges}, where credential sharing is a particularly common scenario with shared hardware devices which support only a single account (though it is possible to use a discrete hardware token that is more easily shared by multiple users with FIDO U2F). On the flip side, credential revocation is another important property desired by end users as previously discussed in Section~\ref{sec:device-bound-credentials}. Lack of an ability to revoke credentials in certain cases has also been a long-standing limitation of SSO~\cite{ghasemisharif2018single}.





\section{Limitations}
\noindent\textbf{Literature Survey:}
Our literature review dataset is thorough but necessarily non-exhaustive, and it is possible our methodology may have missed a small number of papers. For the purposes of this work, our goal is to systematically capture overall trends in academic authentication research, which are reflected in our findings.


\vspace{0.5em}
\noindent\textbf{Understanding E2EE Passkey Synchronization:}
Our understanding and discussion of the security and usability of E2EE synchronization of passkeys relies on the publicly available documentation provided by the vendors.
Hence, it does not allow us to verify that the deployed system behaves as described and the published documents naturally cannot cover all details.
Where necessary we made conservative assumptions about the provided functionalities drawing on knowledge from similar protocols and implementations.

\section{Future Research}
We find four key areas relevant to E2EE authentication and recovery that are not adequately examined in the current literature:

\vspace{0.5em}
\noindent\textbf{FR1:} \textit{Longitudinal recovery studies:}
Account recovery mechanisms are, by definition, more likely to be needed as more time has passed. In 2015, Bonneau et al.~\cite{bonneau2015secrets} found a linear relationship between the time passed since account creation and the proportion of authentication reset requests. Despite this, most usability studies conduct a cross-sectional examination of recovery schemes at a single point in time. A recent longitudinal study from Lassak et al.~\cite{lassak2024comparative} found that for common non-E2EE recovery schemes (email, SMS, recovery questions, and social authentication) the relative convenience of recovery is consistent over time, but similar studies are critical for E2EE services, where the most common E2EE recovery scheme is to require users to maintain a non-human-memorable recovery code. A survey of E2EE email recovery support threads~\cite{holtervennhoff2024mixed} found strong evidence suggesting users are likely to misplace their recovery code over time, motivating further work quantifying this possibility and deriving best practices. Evaluating an optimal balance between mandatory and opt-in periodic backup code confirmation (across a spectrum of backup codes from short PINs to arbitrary pseudorandom recovery codes) is another critical area of future research.

\vspace{0.5em}
\noindent\textbf{FR2:} \textit{Cross-scheme comparative surveys:} We observe that most authentication research analyzes the usability, security, or privacy properties of a distinct category of authentication schemes, with a comparatively number of papers investigating users' perception of a scheme relative to other options. In order for industry to converge on a standard set of authentication schemes, we will need to better understand end-users' perceptions of each scheme relative to other choices. This is especially important in an E2EE context, where providers should offer multiple recovery pathways given that some form of provider-assisted manual recovery is no longer possible.

\vspace{0.5em}
\noindent\textbf{FR3:} \textit{Trusted contact authentication:} Despite being deployed as one of Apple's E2EE cloud storage recovery choices, trusted contact recovery has received comparatively little attention from the academic research community, with just 2.9\% of authentication research since 2012 focusing primarily on this subject as discussed in Section~\ref{sec:lit_survey_results}. The need for a greater focus here is particularly critical in light of recent developments in generative AI, leaving end-users highly vulnerable to impersonation scams and ultimately account compromise, a reality made all the more dangerous given that deployed authentication schemes are increasingly centralized within a single account (e.g., all passwordless authentication credentials may be stored in Apple's cloud storage)

Future work should investigate user preferences in selecting the desired threshold of approval required to regain access, since user preferences will depend on individual situations. Apple uses a single-contact recovery scheme such that only one user is required to approve access, while PreVeil's deployment opts for a threshold 2-of-3 scheme which may be preferred by some users.

\vspace{0.5em}
\noindent\textbf{FR4:} \textit{User perceptions and understanding of E2EE recovery.}
Prior work has shown users have requested recovery schemes are common in everyday web authentication but incompatible with E2EE (e.g., security questions or email/SMS recovery)~\cite{holtervennhoff2024mixed}. Users have also demonstrated poor comprehension of the distinction between account recovery (e.g., the ability to log in) and recovering storage content (e.g., email history, calendar)~\cite{holtervennhoff2024mixed}. Similar patterns of user misconceptions around the significance of the recovery code are seen with mobile cryptocurrency wallets~\cite{voskobojnikov2021u,krombholz2017other}. To date there has been just one academic study of E2EE recovery~\cite{holtervennhoff2024mixed} and more user interface design research is needed to confirm existing findings and to better understand how to improve user understanding around E2EE recovery.

\vspace{-0.6em}
\section{Conclusion}

In this work, we systematize E2EE authentication and recovery mechanisms, identifying numerous unresolved challenges and open areas of research. Given that 62.8\% of top 200 domains now support authentication via passkeys, where the vast majority of passkey implementations store E2EE credential backups, E2EE recovery is a critical are of future research. Each of the various E2EE recovery mechanisms currently offered by cloud services exist somewhere on a spectrum of prioritizing security to prioritizing recoverability. Prior work has suggested the possibility of resolving E2EE-recoverability tradeoffs using biometric authentication in the distant future (in a scenario where the user needs to provide only biometric data to recover account access)~\cite{zinkus2021sok,oduguwa2024passwordless}, but the perennial challenge with biometric and social authentication is keeping pace with ever-more-sophisticated scams that convincingly fake some aspect of human interaction (e.g. voice, live video)~\cite{bethea_voice24,plumb_biometrics24,li2022seeing}. Most importantly, E2EE providers should offer users a greater choice of recovery options to reflect the diversity of users' situations and perceived account value now that E2EE is increasingly targeted at the general public.

\section*{Acknowledgments}
Jenny Blessing is funded by Entrust and Daniel Hugenroth is supported by Nokia Bell Labs.
Ross Anderson made important contributions to the ideas contained in this paper. 
Unfortunately he died on 28th March 2024 before the final version was written; any errors remain our own.

\bibliographystyle{ACM-Reference-Format}
\bibliography{ref}

\appendix

\begin{table*}[h]
\begin{tabular}{@{}lp{5.5in}@{}}
\toprule
\textbf{Authentication Category} & \textbf{Papers}                                                                                                                                                                                                  \\ \midrule
Passwords &
\cite{nisenoff2023two,wash2016understanding,inglesant2010true,wang2016targeted,bonneau2012science,pal2019beyond,das2014tangled,seitz2017differences,wang2023password,ruoti2016strengthening,xu2023improving,al2015impact,li2014large,dunphy2015social,silver2014password,florencio2014password,haque2014applying,wang2023pass2edit,wang2023no,islam2023arana,oesch2020then,pal2022might,ray2021older,becker2018rewards,smith2017augmenting,li2014emperor,pearman2019people,ur2012does,wang2019birthday,lyastani2018better,habib2018user,ruoti2016strengthening,gao2018forgetting,novak2017modeling,segreti2017diversify,melicher2016fast,wheeler2016zxcvbn,koppel2016beliefs,ur2015added,ur2015measuring,theofanos2021passwords,jayakrishnan2020passworld,marne2017learning,lai2017phoenix,alameen_memorability16,hutchinsonmeasuring,alroomi2023measuring,sahin2021don,tan2020practical,seiler2019don,joudaki2018reinforcing,golla2018site,golla2018accuracy,pearman2017let,chatterjee2017typtop,naiakshina2017developers,yang2016empirical,golla2016security,wang2016targeted,huh2015surpass,zhu2014security,mazurek2013measuring,chatterjee2016password,kelley2012guess,blocki2018economics,blocki2023towards,wang2022attack,huaman2021they,ma2014study,pasquini2021improving,sahin2023investigating,gelernter2017password,munyendo2023eighty,liu2023confident,liu2019reasoning,lei2023not,lee2019total,wang2018end,kiesel2017large,blocki2014spaced,de2014very,veras2014semantic,castelluccia2012adaptive,yan2012limitations,abdrabou2022your,oesch2022basically,abdrabou2021think,khamis2019passquerade,naiakshina2019if,huh2017m,meng2016multiple,ur2016users,al2015towards,shay2015spoonful,chowdhury2014passhint,egelman2013does,amft2023would,lee2022password,munyendo2023just,melicher2016usability,shay2014can,blocki2022dalock}                                                                                                                                                           \\ \midrule
Biometric &
\cite{wu2020liveness,he_voiceprint22,li2022seeing,xu2014towards,panjwani2014crowdsourcing,dechand2016empirical,chen2023infinitygauntlet,zhang2017hearing,liu2017vibwrite,sluganovic2016using,zhang2016voicelive,li2015seeing,chen2021real,abdullah2021sok,abdullah2021hear,wu2023depthfake,yan2023spoofing,kassis2023breaking,zhao2020resilience,ba2018abc,li2013unobservable,tian2013kinwrite,tey2013can,fereidooni2023authentisense,yi2023squeez,chen2021user,koushki2021smartphone,lewis2021got,liang2021auth+,van2021understanding,holz2016demand,yang2016free,de2015feel,de2014now,de2012touch,sae2012biometric,jana2020neural,abdrabou2022your}
\\ \midrule

2FA & 
\cite{gilsenan2023security,lee2021security,gerlitz2023adventures,amft2023we,das2018johnny,reynolds2018tale,colnago2018s,abbott2020mandatory,dutson2019don,lassak2024comparative,farke2020you,ulqinaku2021real,gavazzi2023study,mulliner2013sms,lee2020empirical,pfeffer2021usability,redmiles2017you,reese2019usability,sun2015trustotp,reaves2016sending,lei2021insecurity,shirvanian2014two,marky20203d,ciolino2019two,das2018johnny,reynolds2020empirical,turner2023tangible,smith2023if,ghorbani2023systematic,acemyan20182fa,golla2021driving} 
\\ \midrule

SSO &
\cite{balash2022security,somorovsky2012breaking,wang2012signing,ghasemisharif2022towards,yang2018vetting,dietz2014hardening,cho2020will,dimova2023everybody,kroschewski2023save,frederiksen2023attribute,zhang2021passo,ghasemisharif2018single,fett2015spresso,zhou2014ssoscan,mainka2017sok,sun2012devil}
\\ \midrule

Graphical Password &

\cite{meng2016multiple,zhao2013security,al2015impact,uellenbeck2013quantifying,song2017multi,cheon2020gesture,cho2017syspal,abdelrahman2017stay,cheon2023gesturemeter,raptis2021better,hanamsagar2018leveraging,buschek2016snapapp,katsini2018influences,thorpe2014presentation}
\\ \midrule

RBA &
\cite{markert2024understanding,freeman2016you,lin2022phish,wiefling2020more,wu2023him,durey2021fp,senol2024double,markert2022soon,wiefling2019really,wiefling2021privacy,gavazzi2023study}
\\ \midrule
Passwordless & 
\cite{kepkowski2023challenges,lyastani2020fido2,lassak2024aren,lassak2021s,kuchhal2023evaluating,jangid2023extrapolating,troncoso2021bluetooth,feng2021formal,wursching2023fido2,kepkowski2022not,owens2021user}
\\ \midrule
Device PIN & 
\cite{markert2020pin,nicholson2013age,cherapau2015impact,wang2016friend,markert2021security,munyendo2022same,harbach2016anatomy,bace2022privacyscout}
\\ \midrule

Recovery Questions &
\cite{bonneau2015secrets,albayram2016evaluating,hang2015know,zhao2016understanding,hang2015have,micallef2017gamified,albayram2015evaluating,dandapat2015activpass}                             
\\ \midrule
Social Authentication & 
\cite{brainard2006fourth,schechter2009s,kim2012social,polakis2012all,polakis2014faces,javed2014secure,guo2021effect}                                                                            \\ \midrule
Other &
\cite{holtervennhoff2024mixed,keil2022s,harbach2013acceptance}
\\ \bottomrule
\end{tabular}
\caption{\label{tbl:lit_search} Literature search results for authentication and recovery mechanisms. A small number of works appear under multiple categories.}

\end{table*}

\section{Secondary Authentication Mechanisms}
\label{sec:secondary-auth}

\subsubsection{Recovery Questions}
First conceived of in 1990~\cite{zviran1990user}, recovery questions ask users to answer a series of questions based on personal knowledge of the account owner, where the questions are usually determined by the provider but sometimes also user-generated. Frequently used historically, recovery questions are now widely disgraced as a viable authentication scheme after numerous studies showing that answers are low-entropy and often guessable by other individuals personally close to the account owner~\cite{lassak2024comparative,schechter2009s,rabkin2008personal,just2009personal}, a result that has likely only gotten worse as users share troves of personal data online~\cite{pinchot2012s,wired_palin2008}. To make matters worse, studies have also repeatedly found that some users provide untruthful answers as a means of improving their account security~\cite{golla2016analyzing,bonneau2015secrets,lassak2024comparative}, but which has a side effect of making it more likely that the user themselves forgets the correct answer. Bonneau et al.~\cite{bonneau2015secrets} concluded back in 2015 that it is ``next to impossible to find secret questions that are both secure and memorable''.


\hfill \break
\noindent\textbf{Usage-Based Questions:} A suggested variation to reduce the guessability of recovery questions is to use “dynamic” recovery questions in which the answer changes based on account and/or device usage patterns~\cite{albayram2016evaluating,hang2015know,zhao2016understanding,hang2015locked}, as compared with the traditional “static” recovery questions described above. For instance, a service provider may ask questions based on geolocation data~\cite{addas2019geographical,hang2015have}. [Sentence describing general findings.] Users were particularly worried that they wouldn’t be able to recall the correct answers and might lock themselves out of their own account~\cite{albayram2016evaluating}, as user studies found that certain types of data (particularly app installations) are easier to remember than others (e.g., SMS and call history) based on how frequently the data changes.

Authentication based on usage history raises particular concerns for at-risk demographics, such as older individuals who may struggle with memory recall, public figures whose location history is readily accessible, or domestic abuse victims where the attacker is generally familiar with location and communication history.
In summary, we conclude that both static and dynamic recovery questions are unlikely to be suitable as either a standalone authentication mechanism or as a second factor for accounts.

\subsubsection{Email and SMS 2FA}
\label{sec:secondary-auth-mechanisms:sms-2fa}
Email and SMS 2FA are still widely used for recovery today~\cite{li2018email,buttner2023really}, both as the primary 2FA mechanism and as fallback authentication strategies for more secure 2FA schemes (such as backups of 2FA authenticator apps~\cite{gilsenan2023security}). We consider email and SMS 2FA jointly as academic work has shown them to be roughly equally usable in terms of recovery success rates and user perception~\cite{lassak2024comparative,bonneau2015secrets}. SMS has been the most widely deployed 2FA mechanism for at least a decade~\cite{mulliner2013sms,alhusain2021evaluating,gavazzi2023study}, the most common form of which is a time-based one-time password (TOTP). Both code-based and link-based 2FA are vulnerable to social engineering (“real-time phishing”), as users can often be tricked into sharing the code even when told not to do so by the companies. SMS-based authentication codes have well-documented security issues and are easily intercepted via SIM swapping attacks and attacks on the SS7 protocol~\cite{siadati2017mind,lee2020empirical,jover2020security,lee2021security,markert2019view}, but there are nonetheless certain scenarios (e.g., low value accounts, a user posses only a shared email account) that may make SMS 2FA more suitable for an individual use case, and vice versa. SMS 2FA also typically reveals the code on the device lock screen, and in email 2FA some prominent websites (including Google and Facebook) would reveal the code in the email header or preheader~\cite{al2018email}.

\hfill \break
\noindent\textbf{Recovery:} Recovery: While end-users have historically considered email and SMS to be the most usable recovery options, it is nonetheless plausible that users may lose access to one of these factors—for instance, a user may list their university or corporate email as the recovery email for their primary personal email account, and later leave the university or company. Google reported in 2018 that 10\% of users fail email or SMS 2FA~\cite{milka2018anatomy} (e.g., if they no longer have access to the recovery email), though providers attempt to mitigate this by regularly reminding users which recovery options they have set.

\subsubsection{Authenticator App}
\label{sec:secondary-auth-mechanisms:authenticator-app}
Mobile authenticator apps (e.g., Duo Mobile, Microsoft Authenticator) have seen low adoption among the general public~\cite{milka2018anatomy,golla2021driving,petsas2015two} but are frequently mandated in university settings~\cite{abbott2020mandatory,reynolds2020empirical} or by a small number of security-sensitive providers (e.g., Github~\cite{github_2fa}). Academic usability research has found that MFA apps are generally considered easy to use~\cite{colnago2018s,abbott2020mandatory,dutson2019don} but that users perceive the extra step required by MFA as a nuisance~\cite{das2020mfa,marky2022nah,reynolds2020empirical,abbott2020mandatory,de2013comparative}. MFA apps are widely supported among the top domains~\cite{quermann2018state}.


\hfill \break
\noindent\textbf{Recovery from App Loss:} After several years of widespread 2FA app deployment, the academic community has begun to investigate the consequences of app loss (which can frequently occur as a result of device reset, loss, or theft, or because the app backup not including as part of larger device cloud backup).

User concerns over device loss have been a frequent theme with app-based 2FA. Every widely used 2FA app offers different backup and recovery options~\cite{eddy_2fa24}. Several offer cloud backups with iCloud, Google Drive, and sometimes other cloud services, and encrypts backups with a user-chosen password. Others use ``backup codes'' which the user is responsible for storing, and which are easily lost, or offer users the option of backing up to another device via QR code~\cite{gilsenan2023security}.


Prior work has found that recovery is a weak point for 2FA apps that undermines the security mobile app authentication is intended to provide by resorting to the usual array of fallback authentication mechanisms: SMS, email, passwords, and manual recovery~\cite{gilsenan2023security,gerlitz2023adventures}. A spate of recent work has focused on the consequences of losing access to a two-factor authentication (2FA) mechanism, a common scenario for users who use a mobile app for 2FA and a recipe for disaster as mobile devices are regularly lost, changed, damaged, or stolen. Gerlitz et al.~\cite{gerlitz2023adventures} studied how service providers respond to users losing a 2FA mechanism. In 2023, they created accounts at 78 popular websites or mobile apps using 2FA to see what recovery information, if any, was described to the user. They conclude that not enough attention is given to recovery, with 28 of 78 services studied not mentioning anything during the setup phase what backup or recovery procedures, if any, might exist. Amft et al.~\cite{amft2023we} went a step further and analyzed real-world deployments of multi-factor authentication on 71 websites, contacting the sites through public email addresses, support forms as though they were a user who had lost their second factor and going through the manual account recovery process. They found significant variation and inconsistencies among sites, including discrepancies between a given site’s documentation and actual procedures. Accounts on 10 of 71 sites were recovered simply by providing specific knowledge about the account that only the real account owner would know, while 13 sites required some form of official real-world identification, such as a government ID, to regain access. All in all, the authors identify 17 distinct recovery procedures and conclude that they “could not identify best practices regarding MFA recovery procedures” due to the variety.

Gilsenan et al.~\cite{gilsenan2023security} studied backups of two-factor authentication apps that use time-based one-time passwords (e.g., a six-digit code that a user has 30 seconds to provide to a platform), finding commonly shared flaws in backup security, such as that the one-time password data is backed up in plaintext or that SMS is used to authenticate to the backup. The scope of all prior work in 2FA generally assumes two points: (1) that the user recalls or has access to the backup password and only the second factor is lost, and (2) that manual recovery is a possibility since the provider has the ability to reset or remove MFA.


\subsubsection{Hardware Tokens}
\label{sec:the-state-of-web-auth-and-recovery:primary-authenticaion:hardware-tokens}


The well-known weaknesses of authentication flows that rely solely on passwords has motivated the adoption of 2FA technology.
However, SMS-based 2FA (Section~\ref{sec:secondary-auth-mechanisms:sms-2fa}) remains vulnerable to SIM swapping and authenticator apps (Section~\ref{sec:secondary-auth-mechanisms:authenticator-app}) do not protect against live phishing attacks.
Hardware tokens that perform interactive cryptographic protocols with the online service address these issues.

Today, most hardware-based authentication protocols are based on the specifications provided by the FIDO Alliance, which many large companies are part of.
The Universal Second Factor (U2F) protocol~\cite{fido_u2f} allows for interoperable hardware tokens that can provide an authentication proof to different web services.
This additional factor is typically only requested when the user logs in for the first time on a new device.

When registering at a new web service, the client (e.g. the browser) provides the hardware token with origin information that include the domain name.
The hardware token then creates a fresh public-private key pair in its secure memory and derives a key handle based on the origin information.
Both the public key and the key handle are then passed through the client to the web service.

For later authentication, the web service provides the key handle and a challenge to the client which passes it together with the origin information to the hardware token.
The hardware token first verifies that the key handle and origin match.
This prevents other web service from tricking the user to authenticating on a phishing website.
In a second step, the hardware token signs the challenge with the stored private key which the website can verify using the public key.

Hardware tokens typically perform a simple test of user presence by requiring a simple button press on the device.
This ensures that each authentication attempt is known to the user and a malicious app cannot perform authentication requests in the background.
Web services can use the attestation keys provided by the hardware tokens to ensure that the user is using a device that fulfills certain certification standards.~\cite{fido_u2f}

From a usability standpoint, academic work has shown repeated concerns over account lockout upon device loss~\cite{ciolino2019two,reynolds2018tale,das2018johnny,farke2020you}, in addition to general annoyance over the hassle of having to carry and retrieve an additional physical component, which is often reflected in high login timeout or cancellation rates~\cite{ciolino2019two,reynolds2020empirical}. Difficulty of account sharing is another significant concern with hardware tokens~\cite{reynolds2018tale}.

\subsubsection{Risk-Based Authentication}
While not traditionally considered a second-factor since metadata is often unwittingly provided by the user, risk-based authentication (RBA) is a critical strategy providers deploy to distinguish legitimate logins from nefarious access~\cite{wiefling2019really,freeman2016you}. RBA broadly refers to any technical strategy that protects against illicit account access but does not require any information from the user, most commonly asking users to verify their email address~\cite{buttner2023really}. Providers have long resorted to using metadata indicators to determine whether a recovery attempt is legitimate, though high failure rates of 2FA (even basic email and SMS 2FA)~\cite{milka2018anatomy} require companies to carefully balance which logins should be considered suspicious. Broadly, services use an immense variety of mitigation strategies, including throttling (limiting the number of guessing attempts), previously-seen IP addresses, geolocation data, or other metadata available from browser fingerprints~\cite{grassi2020digital,milka2018anatomy,wiefling2019really,freeman2016you,wu2023him,durey2021fp,senol2024double}, and sometimes even monitoring users' behavior post-login~\cite{grosse2012authentication}. Typically, additional verification steps are only deployed if an authentication attempt is deemed suspicious based on the particular metrics and statistical framework used by the account provider.

Both Apple and Google rely heavily on the concept of trusted devices, where logged-in devices are sent a notification to confirm any additional access attempts flagged as suspicious~\cite{risher_blog21,milka2018anatomy}, though this is only relevant when multiple devices are connected to an account. Temporal lockouts, where the user needs to wait a certain period after a correct authentication, are commonly deployed in recovery schemes. For instance, as early as 2008 Gmail only allowed account recovery through recovery questions if the account in question had not been accessed in the past five days~\cite{felten_email2008}. Today, Google deploys temporal lockouts as part of the manual recovery process to ensure a legitimate user has a chance to deny the request~\cite{google_time_lockout}, though research has shown users often fail to recognize malicious sign-in attempts~\cite{markert2024understanding}.

RBA is almost always deployed as an additional authentication measure on top of a pre-existing primary authentication scheme, most commonly in addition to standard password-based authentication~\cite{wiefling2022pump,wiefling2020more}. Prior work has shown it is possible to evade many RBA measures using various cloaking techniques~\cite{andriamilanto2021large,oest2019phishfarm,mirian2019hack,oest2020sunrise,campobasso2020impersonation,lin2022phish}.

\section{Additional Authentication Schemes}
\label{sec:appendix-auth}

\subsubsection{Biometric Authentication}
Finally, a commonly floated approach is for a user to provide some element of their real-world identity to the service provider, such as biometric data. While biometrics are certainly the most resilient form of user authentication, this form of data raises numerous questions around privacy and accuracy. As with recovery questions, biometrics are often used as a component of a multi-factor authentication process, such as unlocking a mobile device with Face ID when a user has also demonstrated physical possession of the device.
However, with regular end-user hardware, we believe are not suitable as a standalone authentication factor for a cloud-based service.

In the last years, new attempts have been undertaken to bridge the gap to allow identifying individuals on a global scale.
One example is the Orb technology that is being deployed as part of the Worldcoin project~\cite{worldcoin}.
The project deploys proprietary hardware and algorithms in order to achieve unique global identification of individuals with very low false-positive and false-negative rates.
One use-case is to provide strong ``proof of personhood'' (and therefore sybill resistance) for Web3 services even where users do not have access to other means identification such as government issued IDs.



\end{document}